%% file: GNN_Learning_topology.tex
\documentclass[%reprint,
arxiv,
 amsmath,amssymb,
]{revtex4-2}

\usepackage{graphicx}% Include figure files
\usepackage{dcolumn}% Align table columns on decimal point
\usepackage{bm}% bold math
\usepackage{easy-todo}
\usepackage{subcaption}
\newcommand{\gnn}{\textsc{Gnn}{}}

\begin{document}

\preprint{APS/123-QED}

\title{Unsupervised Graph Neural Network Reveals the Structure--Dynamics Correlation in Disordered Systems}
\author{Vaibhav Bihani}
\affiliation{
Deparment of Civil Engineering, Indian Institute of Technology Delhi, Hauz Khas, New Delhi, India 110016}
\author{Sahil Manchanda}
\affiliation{
Department of Computer Science and Engineering, Indian Institute of Technology Delhi, Hauz Khas, New Delhi, India 110016}
\author{Sayan Ranu}
\email{sayanranu@iitd.ac.in}
\affiliation{Department of Computer Science and Engineering, Indian Institute of Technology Delhi, Hauz Khas, New Delhi, India 110016}

\affiliation{
Yardi School of Artificial Intelligence, Indian Institute of Technology Delhi, Hauz Khas, New Delhi, India 110016}
\author{N. M. Anoop Krishnan}
\email{krishnan@iitd.ac.in}
\affiliation{Department of Civil Engineering, Indian Institute of Technology Delhi, Hauz Khas, New Delhi, India 110016}

\affiliation{
Yardi School of Artificial Intelligence, Indian Institute of Technology Delhi, Hauz Khas, New Delhi, India 110016}

\date{\today}

\begin{abstract}
\input{abstract}
\end{abstract}

\maketitle

\input{introduction}

\input{Metholodogy}

\input{Experiments}

\bibliography{apssamp}% Produces the bibliography via BibTeX.

\newpage
\include{Appendix}
\end{document}

%% file: abstract.tex
Learning the structure--dynamics correlation in disordered systems is a long-standing problem. Here, we use unsupervised machine learning employing graph neural networks (\gnn) to investigate the local structures in disordered systems. We test our approach on 2D binary A$_{65}$B$_{35}$ LJ glasses and extract structures corresponding to liquid, supercooled and glassy states at different cooling rates. The neighborhood representation of atoms learned by a \gnn{} in an unsupervised fashion, when clustered, reveal local structures with varying potential energies. These clusters exhibit dynamical heterogeneity in the structure in congruence with their local energy landscape. Altogether, the present study shows that unsupervised graph embedding can reveal the structure--dynamics correlation in disordered structures. %The dynamic stability of the obtained clusters is established through a correlation between mean square displacement and average potential energy of clusters.  
%  Our results show a good correlation between the average potential energy of the clusters and the mean square displacement. 
% Our results  show the applicability of graph neural networks in the study of topology of disordered materials.

%% file: introduction.tex
\section{Introduction}
Disordered structures exist in distinct phases such as liquid, supercooled and glassy states~\cite{angell1995formation, debenedetti2001supercooled}. Liquids, when cooled fast enough to avoid crystallization, form glasses~\cite{debenedetti2001supercooled}. While liquids and supercooled liquids are equilibrium states, glasses are considered as metastable state, which continuously relaxes~\cite{debenedetti2001supercooled}. The cooling pathway, such as the rate and the pressure, directly affects the glass structure and dynamics~\cite{PhysRevB.54.15808,li2017cooling,vollmayr1996properties,moynihan1976dependence,smedskjaer2014irreversibility,bhaskar2020cooling}. Thus, although glasses can be considered as ``frozen liquids'', the dynamics of glasses are notably different from that of liquids. However, it is still debated whether the dynamics of these disordered systems are encoded precisely in their local structure.
\looseness=-1

Conventional approaches on understanding the structure-dynamics relationships rely on identifying structural signatures such as medium range crystalline orders (MRCO) that relate to the dynamical heterogeneity in disordered systems~\cite{watanabe2008direct,kawasaki2007correlation,tah2018glass}. However, attempts to predict the dynamical heterogeniety directly from the structure have been barely successful, although indirect correlations have been established~\cite{coslovich2006there,widmer2004reproducible,widmer2006predicting,widmer2007study,cubuk2015identifying,malins2013lifetimes}. Recently, data-driven approaches employing machine learning (ML) was used to predict atomic dynamics directly from the structure. Specifically, ``softness''~\cite{schoenholz2016structural,doi:10.1073/pnas.1610204114}, a machine learned parameter, was introduced to predict the tendency of atoms to rearrange purely based on their local structure in a disordered system. Further, Bapst et al.~\cite{bapst2020unveiling} used a graph neural network (\gnn) to learn the structural descriptors capable of predicting propensity~\cite{berthier2007structure,widmer2007study} of atoms to move under shear stress. Although both these works~\cite{schoenholz2016structural,bapst2020unveiling} confirmed that the dynamical heterogeneity could be predicted directly from the structure, they employed a supervised ML approach~\cite{lecun2015deep,carleo2019machine}, where the models were trained explicitly to learn the descriptors that predict the dynamics of atoms. In other words, the data on both structure and dynamics was used to learn their interrelationship through supervised ML, which was then used to model unseen systems. This infuses an inherent bias to learn the features that best represent the individual dynamics of atoms. Hence, the descriptors may not be suitable for other tasks such as predicting the collective dynamics such as self-organization or even learning the local motifs in glass structures~\cite{jones2020glassy,boolchand2005self}. Further, training the model require a priori knowledge of both the structure and the dynamics, which may not be available for many systems. 

Here, using \gnn{s}, we learn the structural features of disordered systems directly in an unsupervised fashion~\cite{hamilton2017inductive}. Specifically, by considering disordered structures as graphs~\cite{trudeau2013introduction}, we allow the model to learn local atomic environment in an unsupervised fashion. We show that the \gnn{} is able to identify the features that, in turn, discovers local clusters in the disordered structures reminiscent of the MRCO. We test our approach on a well established 2D binary LJ glass of composition A$_{65}$B$_{35}$ to form glasses from the melt at different cooling rates ranging over four orders of magnitude. The clusters discovered by the \gnn{} exhibit a good correlation in terms of their dynamics and energy, revealing the existence of a structure--dynamics correlation.

%% file: Metholodogy.tex
\section{Methodology}

\begin{figure}[h]
\hspace{-0.2in}
\includegraphics[width=\linewidth]{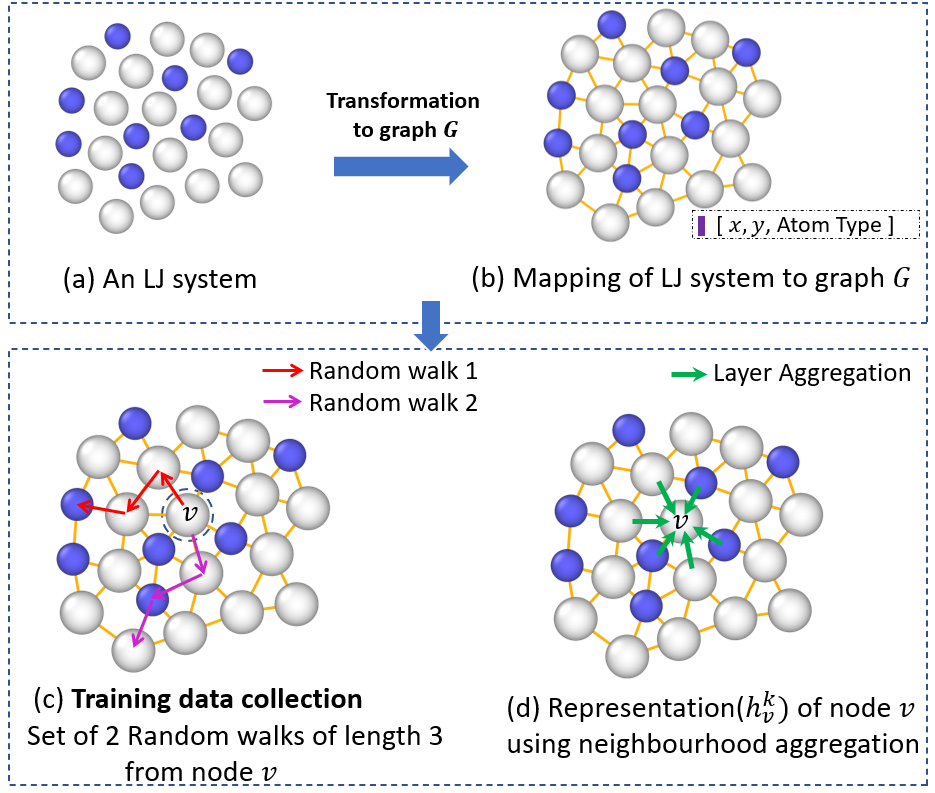}
\caption{Proposed framework for learning the structural representation of disordered systems.}
\label{fig:arc}
\end{figure}

To learn the local topology of disordered systems, we use \gnn{s}. Figs.~\ref{fig:arc}(a)-(b) describe the transformation from disordered structure to a graph. First, we transform each of the disordered structures as an undirected graph~\cite{trudeau2013introduction} $G=(V,E)$. The node set $V$ and edge set $E$ represent the atoms and the set of bonds between these atoms, respectively. Here, the bonds of the atoms are with their first neighbors with the cutoff distance as the first minima of the partial pair distribution functions (PDFs). Thus, $\mathcal{N}(v)$ represents the set of nodes that are neighbors to the node $v$ in graph $G$. Further, each node $v \in V$ consists of the following node attributes ($h_v^{0}$): \textbf{(1)} one-hot encoding representing the atom (node) type (\textit{i.e.}, type \textbf{A} or \textbf{B}). \textbf{(2)} Coordinates of the node $v$ in two dimensions $(x_v,y_v)$. 

Our goal is to understand the local structure of such disordered systems. Towards this end, we utilize \textsc{GraphSage}~\cite{hamilton2017inductive} to learn the node representation in a way that captures topological structure of each node's neighborhood (see Fig.~\ref{fig:arc}(c),(d)). Hence, for each node $v \in V$, we learn how to aggregate information from its neighboring nodes $\mathcal{N}(v)$ as
\begin{equation}
\label{eq:GNN}
\mathbf{h}_{v}^{k} \leftarrow F\left(\mathbf{W^k} \cdot \operatorname{MEAN}\left(\left\{\mathbf{h}_{v}^{k-1}\right\} \cup\left\{\mathbf{h}_{u}^{k-1}, \forall u \in \mathcal{N}(v)\right\}\right)\right.
\end{equation}
where $h_v^{0}$ represents the raw feature information of nodes comprising of the atom type and coordinates, $h_v^{k}$ represents the node information after $k-$rounds of message passing with the neighbors, and $F$  represents an activation function which in our case is the sigmoid function. 

The purpose of the \gnn{} is to construct representations for each node integrating both the nodes' initial raw feature information and the information about the local graph structure that surrounds them. Specifically, each node aggregates its own information and information from its $k{-}hop$ neighborhood to update its representation, i.e.,  $h_v^{1}... h_v^{k}$ where $k$ denotes the depth of the \gnn{} and $h_v^{k}$ denotes the final representation of the node $v$. The term $\mathbf{W}^{k}$ is a learnable parameter of the \gnn{} at depth $k$. In the case of a disordered atomic structure, $\mathbf{W}^{k}$ learns to aggregate feature information from neighboring atoms and pass this information to generate an improved representation for the central atom $v$ incorporating the local neighborhood features. With a higher depth $k$, nodes incrementally gain more and more information from further reaches of the graph. For instance, $k=m$ aggregates information from the $m^{th}$ hop neighborhood of the node $v$. However, higher values of $k$ in \gnn{}s, such as $k > 3$ could lead to poor performance~\cite{liu2020towards}.

Next, we train the \gnn{} by minimizing an objective function.   
Since our aim is to learn the local features of the structure without any additional information, we resort to unsupervised learning. In order to train the model, we generate a fixed-length ($L$) random walk~\cite{aldous1995reversible, hamilton2017inductive} from each node $v$ in the graph, which visits different nodes in $v$'s neighborhood. This random walk allows the exploration of the local structure surrounding node $v$. Fig.~\ref{fig:arc}(c) depicts 2 random walks of length 3 from a node $v$.  
Note that in contrast to previous approaches, here we are not forcing the \gnn{} to learn any task-specific embedding such as predicting the dynamics. The embeddings are purely a representation of the local neighborhood of a node and hence is purely structural in origin. In other words, the representation learned by \gnn{} is purely based on the atomic structure and has no bias toward their dynamics.
\looseness=-1

Through these random walks, we identify nodes $u\in V$ that co-occur in $v$'s neighborhood. Without loss of generality, we consider these node pairs $(v,u)$ as \textit{positive samples}, that is representing the neighborhood of $v$. Node pairs $(v,u_n)$ that do not co-occur in random walks are considered as \textit{negative samples}, which is representative of a random node absent from $v$'s neighborhood. Next, our unsupervised objective function given by Eq.~\ref{gnn:loss} is optimized to learn the parameters of the Graph neural network.
\looseness=-1
\begin{equation}
\label{gnn:loss}
\mathcal{L}\left(z_{u}\right) = -\log \left(\sigma\left(z_{v}{ }^{T} z_{u}\right)\right)-Q \cdot E_{u_{n} \sim p_{n}(u)} \log \left(\sigma\left(z_{v}{ }^{T} z_{u_{n}}\right)\right).
\end{equation}
Here, $z_v=h_v^k$, where $k$ is the number of layers of the \gnn{} and $Q$ determines the number of negative samples from the negative distribution $p_{n}(u)$ (consisting negative samples). The minimization of the objective function encourages nodes that are in same neighborhood (\textit{positive samples}) to have similar representation and vice-versa. As evident from  Eq.~\ref{gnn:loss}, the  dot-product (hence, similarity) of representations of positive node pairs $(v,u)$ that co-occur in random walks is encouraged to be higher, while the opposite is true for the representation of negative node pairs $(v,u_n)$.

%% file: Experiments.tex
\section{Results and Discussion}

We test our approach on a 2D binary LJ system. We perform MD simulation of melt-quenching LJ system~\cite{bruning2008glass} of 500 atoms with composition A$_{65}$B$_{35}$ to prepare glasses at different cooling rates. All the simulations are carried out using the LAMMPS package~\cite{LAMMPS}. The interaction between the particles is governed by
\begin{equation}
\label{eq:lj}
V_{\mathrm{LJ}}(r)=4 \varepsilon\left[\left(\frac{\sigma}{r}\right)^{12}-\left(\frac{\sigma}{r}\right)^{6}\right]
\end{equation}
where $r$ refers to the distance between two particles, $\sigma$ is the distance at which inter-particle potential energy is zero and $\varepsilon$ refers to the depth of the potential well. Here, we use the LJ parameters $\varepsilon_{AA}=1.0$, $\varepsilon_{AB} = 1.5$, $\varepsilon_{BB}= 0.5$, $\sigma_{AA} = 1.0$, $\sigma_{AB}= 0.8$ and $\sigma_{BB}= 0.88$. The mass for all particles is set to $1.0$. All the quantities are expressed in reduced units with respect to $\sigma_{AA}$,  $\varepsilon_{AA}$, and Boltzmann constant $k_B$. We set the interaction cutoff $r_c=2.5\sigma$~\cite{singh2013ultrastable} and the time step $dt=0.003$ for simulations. 

We perform all the simulations at constant volume. For preparing the glass structure, first the ensemble is equilibrated at high temperature $T=2.0$ where it equilibrates in the liquid state. Next, it is cooled down to low temperature of $T=0.05$, at $4$ different cooling rates ranging from $3.33\times10^{-3}$ to $3.33\times10^{-6}$. The cooling is carried out at NVT, using Nose-Hoover thermostat in temperature steps of $0.05$. At each of these steps, the atomic structure of the system along with their density and enthalpy is noted. In order to avoid the noise due to thermal fluctuations, the glass structures are minimized at 0 atm pressure to obtain the ground state enthalpy and density~\cite{singh2013ultrastable}. Accordingly, we obtain $160$ minimized glass structures, $40$ from each cooling rate, representing the ground state structures at different temperatures and states. 
\looseness=-1

To train the \gnn{}, we set $k = 2$, representing the information aggregation from up to the $2^{nd}$ neighbors. We observe that higher values of $k$ could accumulate noise, which is analogous to the PDF in disordered systems which at larger distances saturates to 1. The choice of $k=2$ also resulted in learning better representations as evidenced by the performance on the validation set as discussed in Fig.~\ref{fig:Layers_test} in section S4 in the Appendix. For performing random walk, we set $L=4$ and $Q=10$ and number of random walks starting from each node to be $10$. Effect of random walk length parameter is shown in section S7 of Appendix. We set the size of node representation to be $64$. Finally, we use the training graphs to optimize the parameters of \gnn{}. Using the learnt parameters of \gnn{}, we generate node representations for graphs that were not seen during the training phase. The parameters used for training \gnn{} are specified in detail in section S4 in Appendix.

\begin{figure}[h!]
\includegraphics[width=\linewidth]{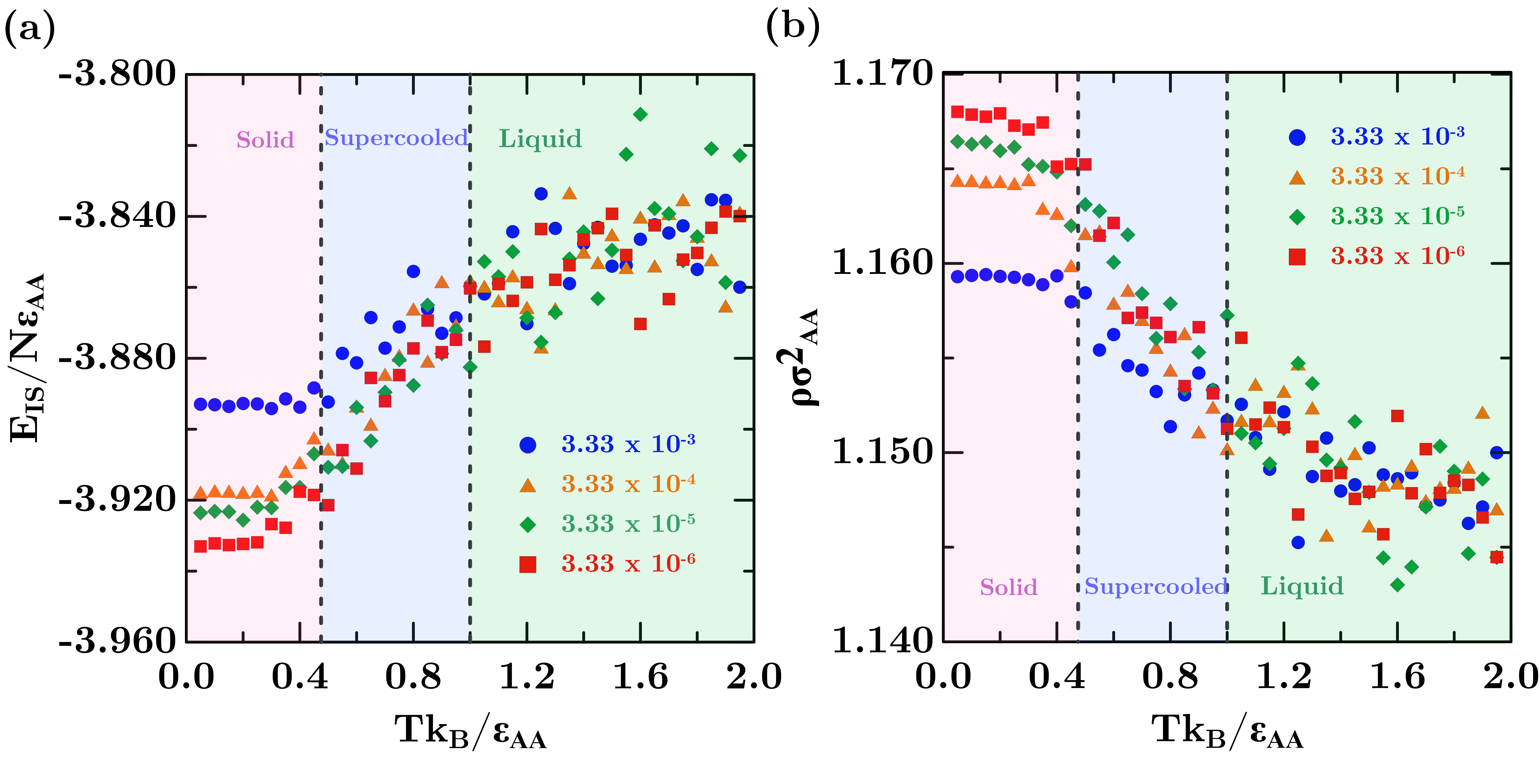}
\caption{ Variation in \textbf{(a)} potential energy, and \textbf{(b)} density with respect to temperature for different cooling rates of the binary LJ system. The liquid, supercooled liquid, and the solid (glassy) regions are highlighted.}
\label{fig:cooling_curve}
\end{figure}

\begin{figure}[h!]
\includegraphics[width=\linewidth]{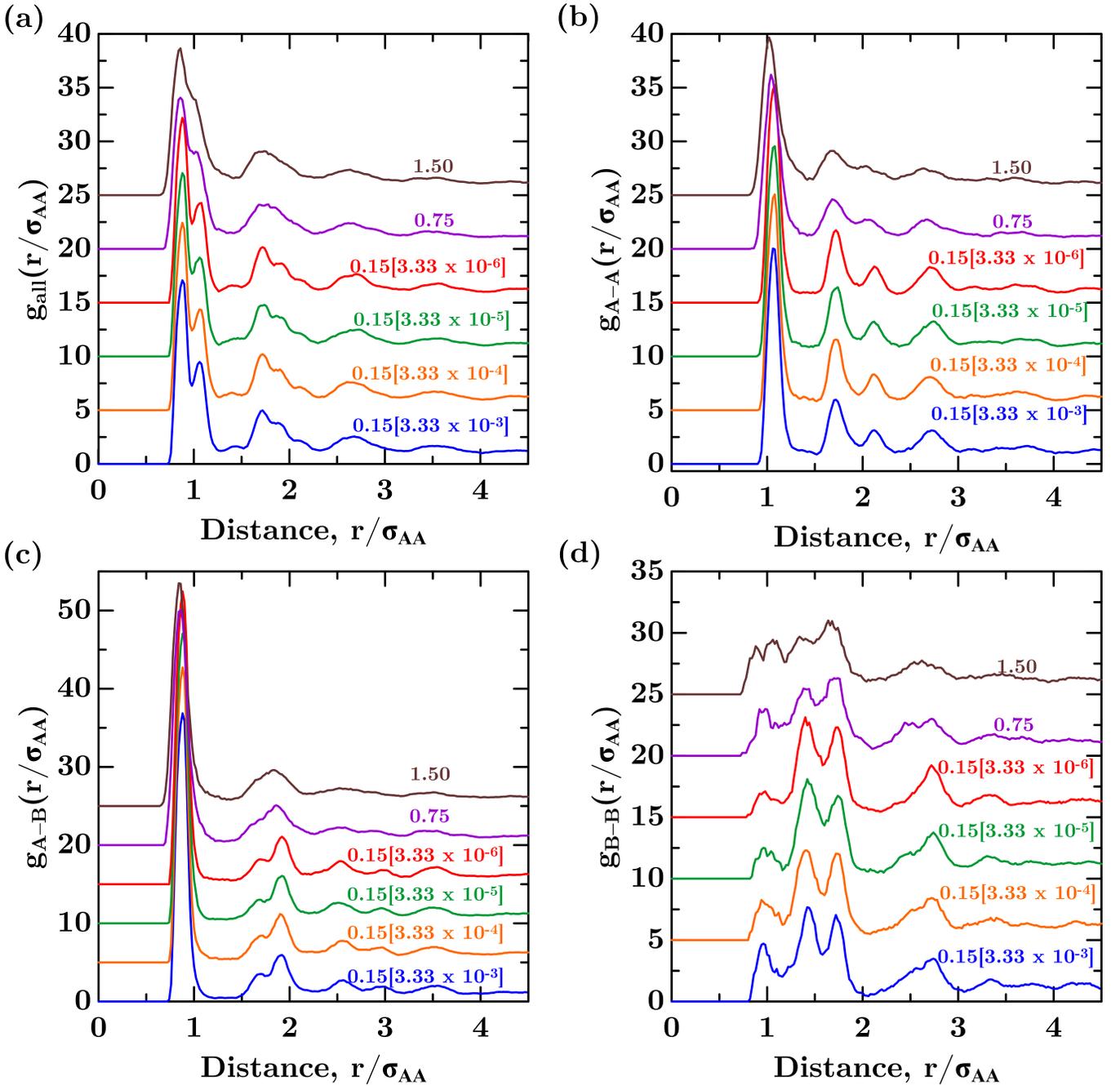}
\caption{(\textbf{(a)} Total, (b) A-A, (c) A-B, and (d) B-B partial pair distribution functions for the binary LJ system in the liquid, supercooled liquid and glassy states. An offset in the $y$-direction is provided for different systems to improve clarity in representation.}
\label{fig:pair_dist}
\end{figure}

First, we analyse the change in the enthalpy and density of the disordered structures, when they are cooled at different rates.  Figs.~\ref{fig:cooling_curve}(a) and (b) represent the variation in enthalpy and density of the structures upon quenching. In agreement with theoretical understanding~\cite{singh2013ultrastable,vollmayr1996properties,li2017cooling,bhaskar2020cooling}, we observe that lower cooling rates lead to more stable glasses with lower enthalpy and higher densities. We also observe that the density and enthalpy of the structures in the liquid ($T > 1.0$) and supercooled ($0.6 < T < 1.0$) regimes are indistinguishable, despite the difference in their cooling rates---thanks to their equilibrium nature~\cite{debenedetti2001supercooled}. In contrast, the solid (glassy, $T < \sim 0.6$) state exhibits clear variation for the structures obtained with different cooling rates. Note that the enthalpy of a system have major contributions from the short-range interactions in a structure, while the density could be attributed to the medium- and long-range order of the system. Accordingly, we observe that the disordered structures simulated have notable differences both in terms of their short-range interactions and the long-range order.  

Next, in Fig.~\ref{fig:pair_dist} we compare the total and partial PDFs of the structures obtained at different temperatures and cooling rates. Since liquid and supercooled structures does not show any significant difference of cooling rate (see Fig.~\ref{fig:cooling_curve}), only one PDF is plotted representing them. We observe that despite the denser and more stable structure (in terms of enthalpy) of glass obtained at lower cooling rate, the total PDFs obtained are almost identical. Except a minor broadening of the first peak of B-B partial PDFs, the partial PDFs of different glassy structures are also comparable. This results shows that merely distance distribution cannot capture the effect of cooling rate on the local structure. Note that extensive studies have been carried out to understand the local structure of LJ glasses with different cooling rates and readers are referred to these works for further differences and similarities on the local structure with respect to cooling rate~\cite{vollmayr1996properties}.
\looseness=-1

Towards our goal of gaining better understanding of the different local regions in the disordered structure (graph), we use the unsupervised graph embeddings learned by the \gnn{}. Specifically, we cluster the nodes of a graph based on their node representations using a density-based clustering algorithm OPTICS~\cite{ankerst2008ordering} and cosine similarity as our distance function between two nodes. 

Specifically, nodes (atoms) in the graph are ordered such that spatially closest points become neighbors in the ordering. Then based upon a distance parameter, representing the density, a decision is made to  accept/reject points to be part of a cluster. Details of the OPTICS clustering algorithms and hyper-parameters used are given in section S2,S3 and S5 in  Appendix.

\begin{figure}[h!]
\includegraphics[width=\linewidth]{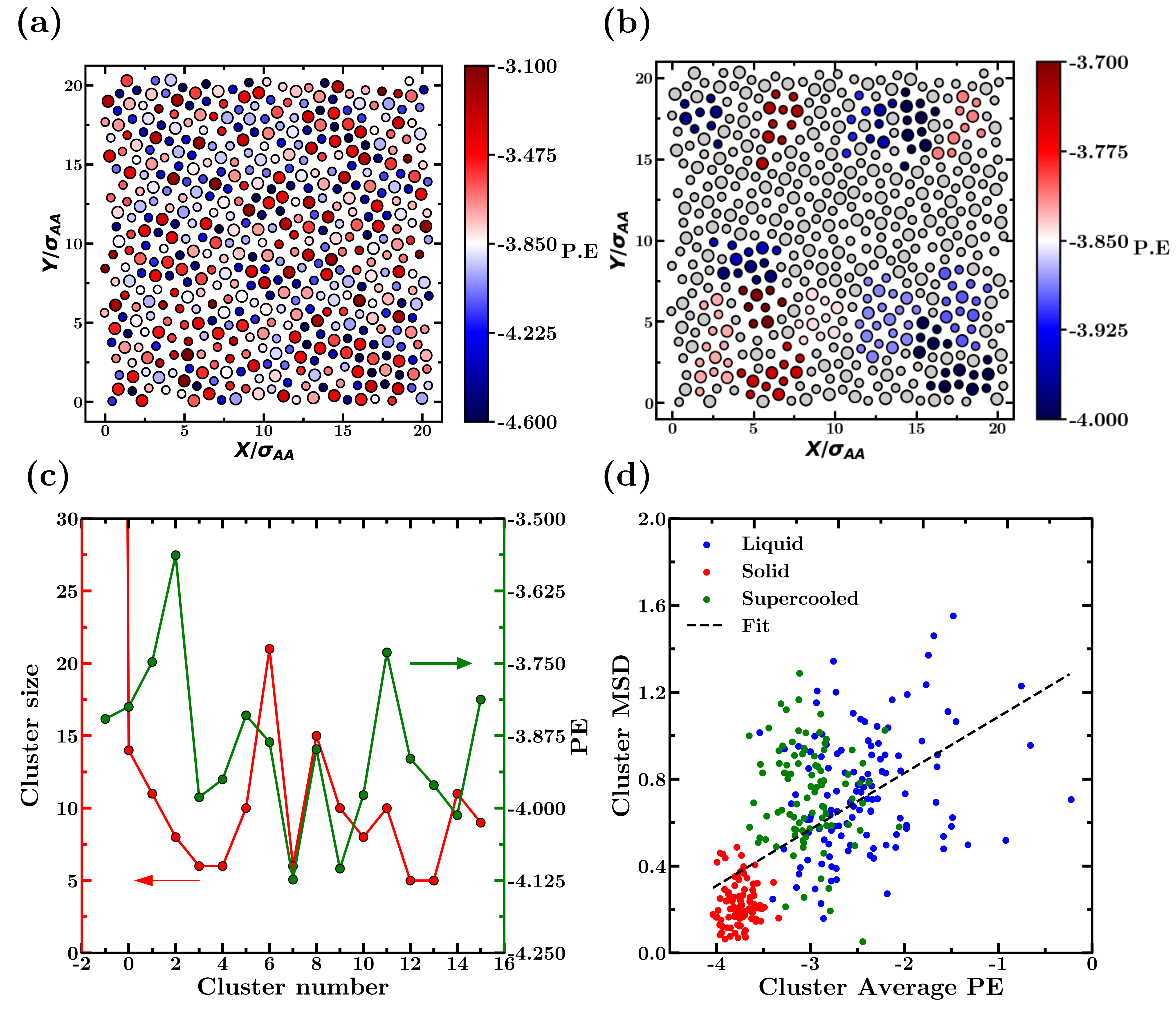}
\caption{(a) Potential energy per atom, (b) average potential energy of the clusters obtained. (c) Cluster size and potential energy of each obtained cluster. (d) Average potential energy of the cluster with respect to the MSD.}
\label{fig:cluster}
\end{figure}

Now, we analyze the LJ system when atoms are clustered based upon the approach described above. Fig.~\ref{fig:cluster}(a) shows the potential energy per atom for an arbitrary glass structure from the simulations. We observe that the energy distribution seems to be random with no local order. In contrast,  Fig.~\ref{fig:cluster}(b) shows the average potential energy of the clusters obtained from OPTICS clustering of node representations generated by the trained \gnn{}. Note that nodes colored in grey are not part of any cluster. i.e., they are outliers. Additionally, we also show the reachability distance plot generated by OPTICS in section S6 in Appendix. Fig.~\ref{fig:cluster}(c) shows the cluster size and average potential energy of the clusters. We clearly observe several high and low energy clusters in the structure.  It can be observed from the limits of color bar in Fig.~\ref{fig:cluster}(b) that the variation in average potential energies of the obtained clusters, although small in comparison to that of individual atoms, is of higher range than that observed in the cooling curves of Fig.~\ref{fig:cooling_curve}. Interestingly, the local clusters observed are reminiscent of the MRCO observed in supercooled and glassy systems based on orientational order parameters~\cite{watanabe2008direct,kawasaki2007correlation,tah2018glass}. This suggests that the unsupervised embeddings learned by the \gnn{} is able to capture the local structural order (and disorder) in these systems.

In order to analyze whether the learnt cluster correlates to the dynamical heterogenity observed in these systems, we study the curvature of the local energy landscape of the clusters following an established methodology ~\cite{PhysRevX.7.031019,debenedetti2001supercooled}. First, we perform multiple energy minimizations to ensure that the samples are at their ground state. Then, we provide a sudden energy bump to the system corresponding to a temperature of $T = 0.25$. The temperature is chosen to be high enough to allow potential motion between low-energy barriers but low enough to avoid glass transition or melting of the system. Finally, we allow the system to evolve in the micro-canonical ensemble (NVE) for 100 thousand steps. The MSD $(r^{2})_C$ of a cluster $C$ is obtained as
\begin{equation}
\label{eq:MSD}
(r^{2})_C= \left[\frac{1}{N_C}\sum_{i=1} ^{N_C}(\vec{r_{i}}-\vec{r_{i,0}})^{2}\right]
\end{equation} 
where $N_C$ represents the number of atoms in a cluster, $r_{i}$ are the positions of each atom after 100 thousand steps of dynamics following the energy bump, and $r_{i,0}$ are the positions in the inherent structure. 

Stable low-energy structures represented by narrow and deep local energy landscape exhibit low MSD, while high energy structures, exhibiting a shallow and broad landscape, exhibit high MSD. To understand the correlation between the clusters identified by \gnn{} and dynamical heterogeniety, we plot the  MSD of the clusters with respect to the average potential energy of the clusters (see Fig.~\ref{fig:cluster}(d)). Note that each data point in the plot corresponds to a cluster, and the color represent the state (liquid, supercooled, solid) from which it was obtained. We observe a direct positive correlation with the cluster potential energy and the MSD with a Pearson coefficient of 0.747, representing high statistical significance. The variations in the cluster MSD even within a given disordered structure~(see Fig.~\ref{fig:MSD_struct} in section S1 of Appendix), in congruence with the local potential energy, suggests the existence of dynamical heterogeneity in these systems. These heterogenieties are captured effectively by the graph embeddings.

\section{Conclusion}
Altogether, present study reveals a novel method, employing unsupervised \gnn{}, for learning the local structure in disordered systems. We demonstrate that these local structures exhibit a direct correlation with the dynamical heterogenieties present in the disordered systems. It is worth noting that in contrast to previous approaches, where ML was trained against a data to predict dynamics from structure, the present approach reveal local structural motifs in an unbiased manner purely from the structure. Further analysis of these clusters could potentially reveal hidden order within a disordered structure, that, in turn, govern their dynamics. In addition, the learned embedding could further be potentially used as a representation or a reduced-order feature of the disordered system to predict their properties. Our approach thus paves way for gaining an understanding on the structure--dynamics correlation in disordered systems that are not achieved by conventional methods.

The code is available at :~\url{https://github.com/M3RG-IITD/ML-DOS}

%% file: Appendix.tex
\begin{center}
    \Large \textbf{Supplementary Material}
\end{center}

\section*{S1. Cluster mean squared displacement in 2D LJ System}

 \begin{figure}[h!]
\includegraphics[width=\linewidth]{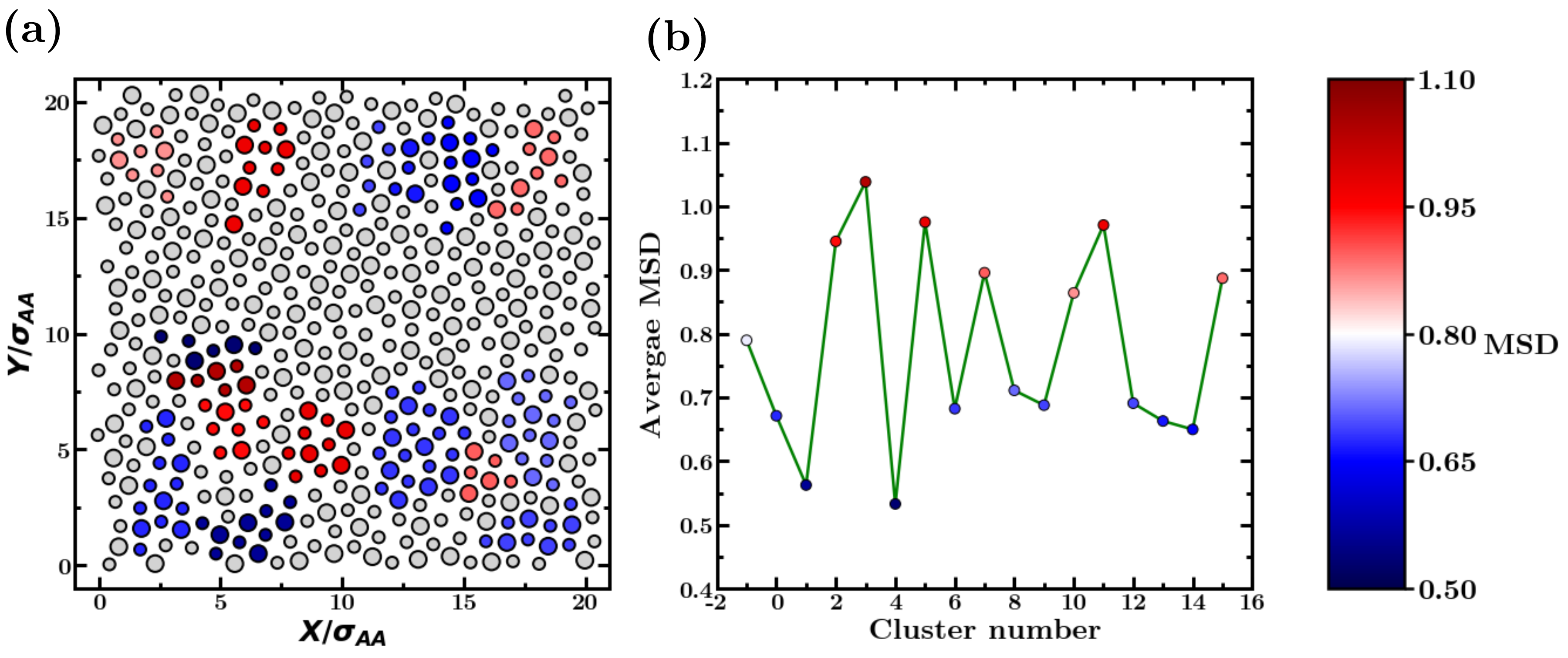}
\caption{(a)Cluster mean squared displacement in 2D LJ system, (b) Cluster mean squared displacement vs Cluster number }
\label{fig:MSD_struct}
\end{figure}

In Fig.~\ref{fig:MSD_struct}(a) and (b) we show the obtained mean squared displacement of clusters in our 2D LJ system. Clustering the graph embeddings thus captures the existence of dynamic heterogeneity in these disordered systems. 

\section*{S2. Density based clustering:}
\begin{figure}[h!]
\includegraphics[width=\linewidth]{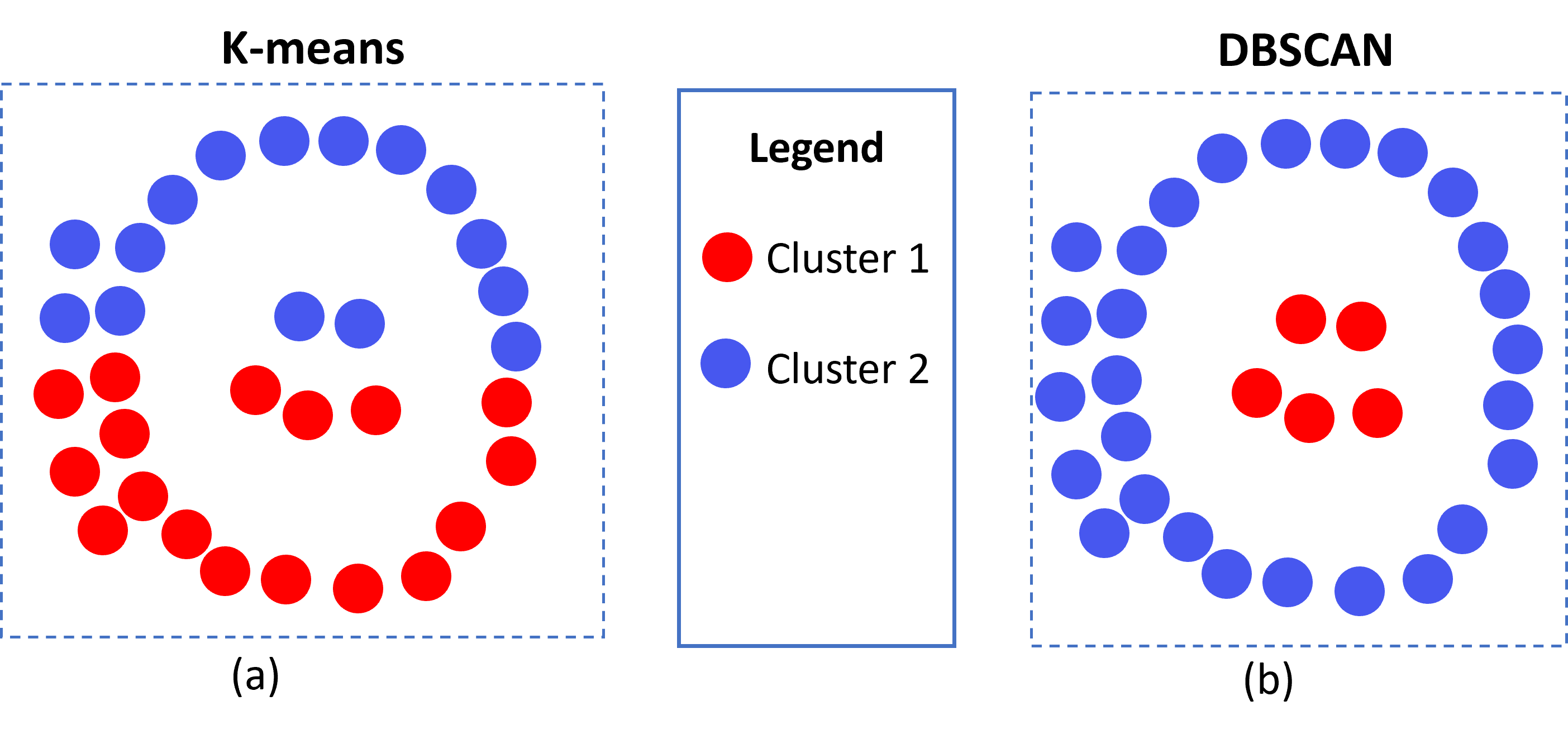}
\caption{An example of DBSCAN vs centroid based clustering (K-means) for points in 2D space.}
\label{fig:dbs-kmean}
\end{figure}

Density based clustering identifies groups/clusters in data based upon the intuition that each cluster is a contiguous region of high point density, separated from other clusters by regions of low point density.  Fig.~\ref{fig:dbs-kmean} shows an example of density based clustering obtained using the DBSCAN algorithm~\cite{ester1996density}. We can clearly see that DBSCAN generated clusters where density of points is high and is  separated by regions of low density of points. Density based clustering can generate clusters of any shape, which has advantage of other methods such as the popular \textit{centroid-based} technique \textit{K-means}, which generates clusters that are more or less spherical.

\begin{figure}[h!]
\includegraphics[width=\linewidth]{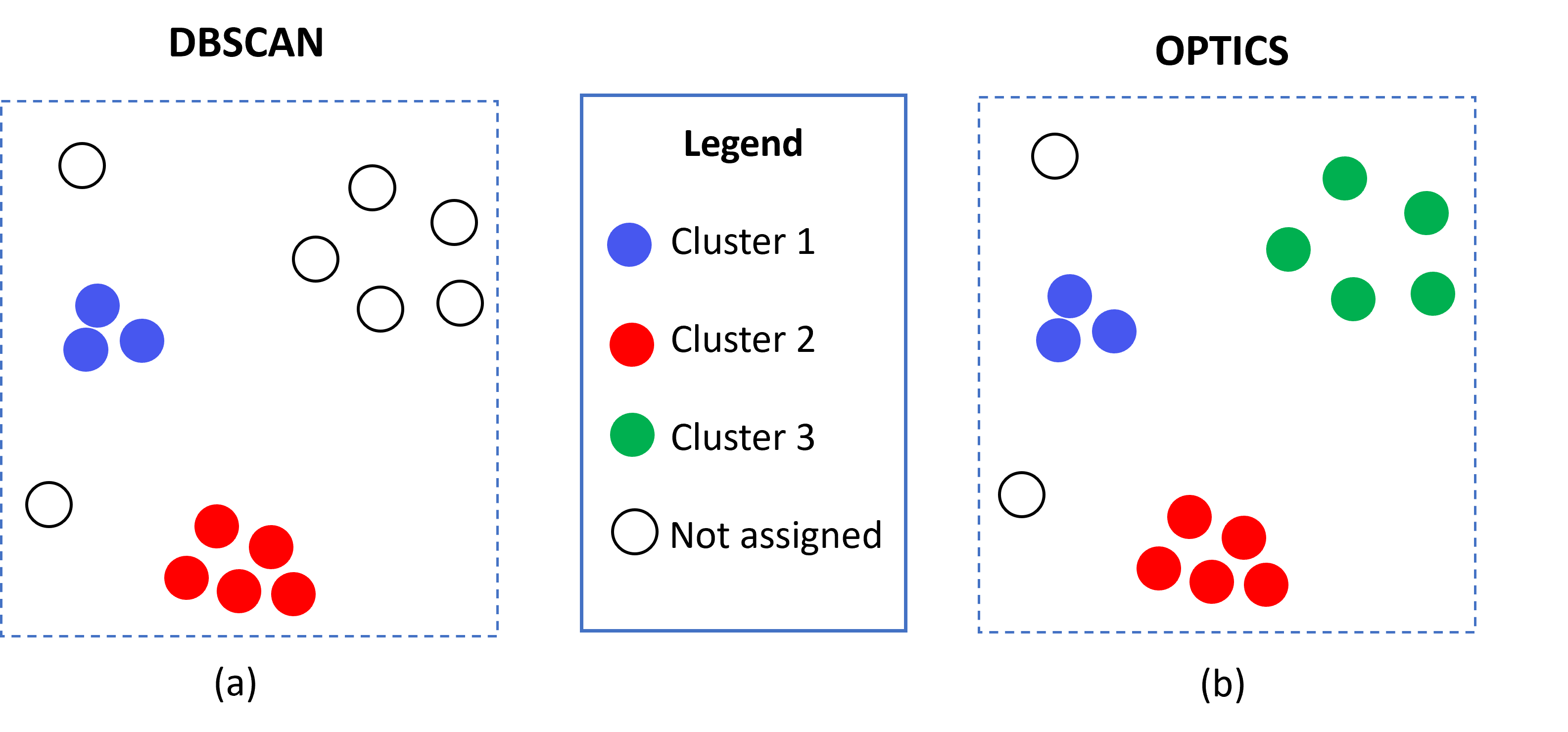}
\caption{An example of DBSCAN vs OPTICS for points in 2D space.}
\label{fig:dbs-opt}
\end{figure}

    However, DBSCAN fails to identify clusters when data has clusters of varying density.  As an example, let us consider Fig.~\ref{fig:dbs-opt}(a), which consists clusters of varying densities. DBSCAN fails to identify one of the clusters. The prime reason responsible for this behavior is that DBSCAN uses a fixed cutoff distance between two points to consider them as neighbors. However, a \textit{larger} distance in one cluster could be a \textit{shorter} distance in another cluster as can be seen in Fig.~\ref{fig:dbs-opt}(a). Further, a solution that uses larger cutoff could lead to merging of multiple clusters and a lower cutoff could lead to missing out points to be part of a cluster as in Fig.~\ref{fig:dbs-opt}(a). Hence, finding the optimal cutoff is non-trivial.
    \looseness=-1

To overcome this limitation, the OPTICS  (Ordering Points To Identify the Clustering Structure)~\cite{ankerst1999optics} algorithm was proposed. Specifically, instead of relying on a fixed cutoff distance, points are ordered linearly such that spatially closest points become neighbors in the ordering. This ordering is then utilized to perform clustering.  As we can see in Fig.~\ref{fig:dbs-opt}(b), using OPTICS, we are able to identify regions having varying densities. 

\section*{S3. Optics Clustering Algorithm}
\label{app:optics}
As discussed above,  OPTICS~\cite{ankerst1999optics} is a density-based clustering algorithm, like DBSCAN, with the additional advantage of being able to identify clusters of varying densities.
 We next introduce few definitions for better understanding of the algorithm. Please refer to Fig.~\ref{fig:core1} for a visual depiction. 
\begin{itemize}
\item $\epsilon$ parameter: Two points are considered neighbors if their distance is less than $\epsilon$. In our case it translates to the cosine distance between the learned representations of two atoms (nodes). Specifically for our case, without loss of generality, an atom $p$ is neighbor of an atom $q$ if $CosineDistance(h_p^k,h_q^k) < \epsilon$. Here $h_p^k$ refers to the representation of the $p^{th}$ atom at  $k^{th}$ layer of the \gnn{}.

\begin{figure}
\centering
\begin{subfigure}[b]{0.6\textwidth}
\centering
   \includegraphics[width=0.8\linewidth]{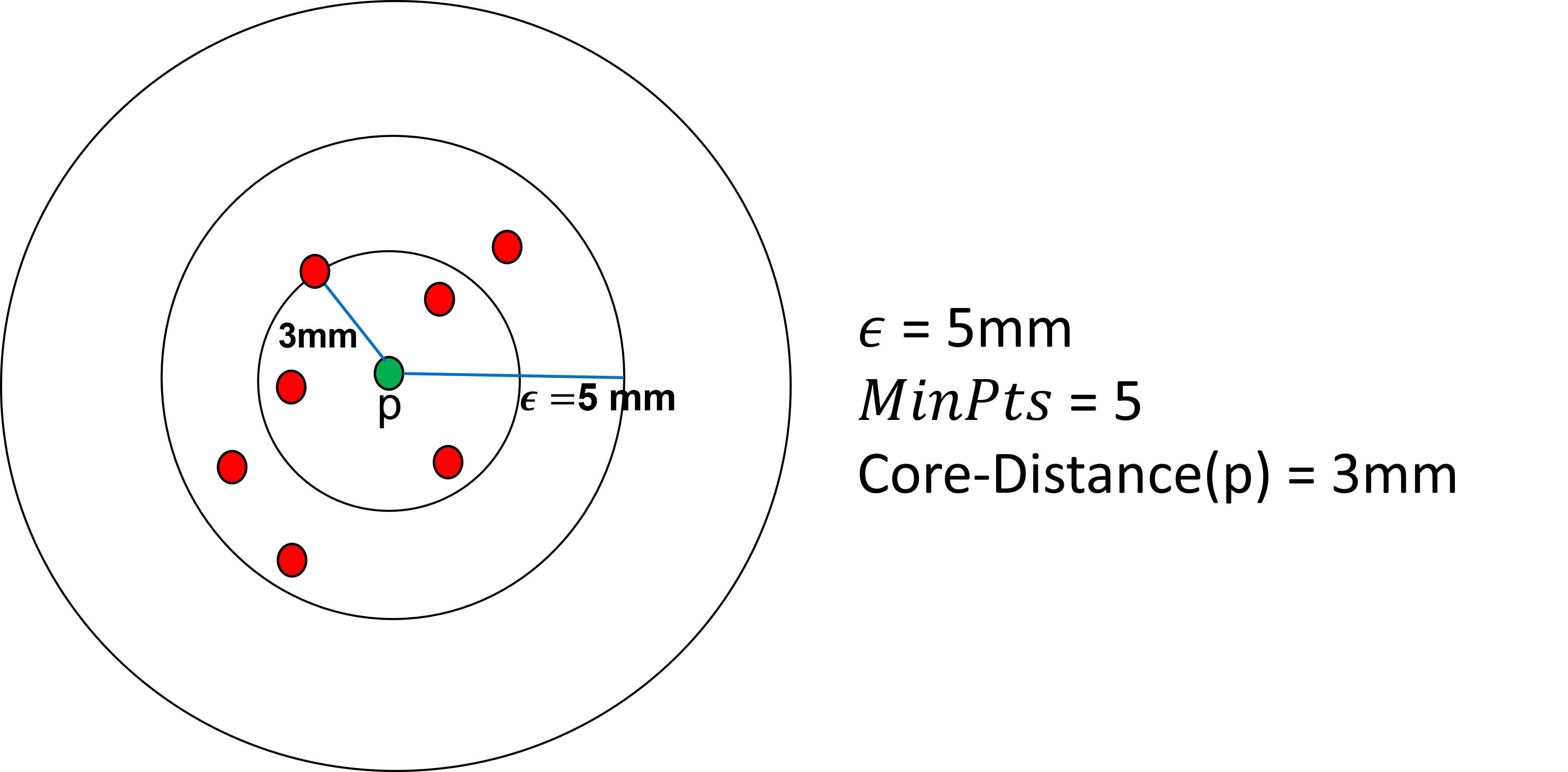}
   \caption{Core point }
   \label{fig:core1} 
\end{subfigure}

\begin{subfigure}[b]{0.55\textwidth}
   \centering
   \includegraphics[width=0.75\linewidth]{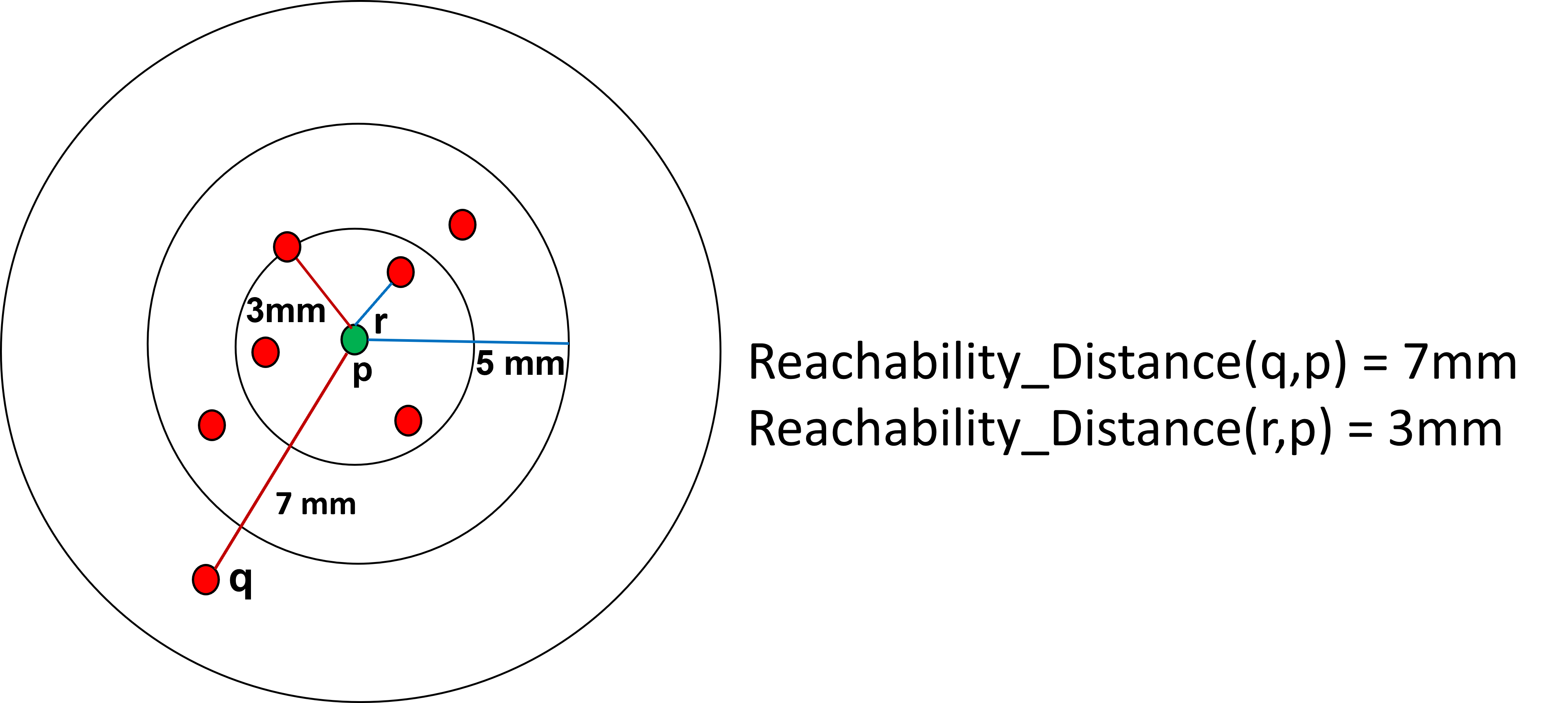}
   \caption{Reachability distances}
   \label{fig:core2}
\end{subfigure}
\caption{OPTICS: Core points and reachability distance of points}
\end{figure}

\item $\epsilon$-Neighborhood  $N_\epsilon({p})$: The points that are within $\epsilon$ distance of a node $p$. 

\item $MinPts$: Minimum number of points that must be within  $\epsilon$ distance of an atom $p$ in order to form a cluster.

\item \textbf{Core Point:} A point is considered a core point if it has more than $MinPts$ points within $\epsilon$ distance.

\item \textbf{Core Distance:} Minimum $\epsilon$ required to make a point $p$ a core point.

{\small
\begin{alignat}{2}
\nonumber
&\text{core-dist}_{\epsilon , \text{\textit{MinPts}}}(p)=\\ 
&\begin{cases}
    undefined, \; \; \quad \quad \quad \quad \quad \quad \text{if } |N_{\epsilon}(p)| < \text{\textit{MinPts}}  \\
    \text{\textit{MinPts}-th smallest distance in } N_{\epsilon}, \;  \text{else}
\end{cases}
\label{eq:core}
\end{alignat}}

\item \textbf{Reachability Distance:} Reachability distance between a point $p$ and $o$ is the maximum of the \textit{Core Distance} of $p$ and the  distance between $p$ and $o$. For point  $o$ that is reachable from a point $p$:

\begin{small}
\begin{alignat}{2}
\small
\nonumber
&\text{reach-dist}_{\epsilon , \textit{MinPts}}(o, p){=}\\
&\begin{cases}
\text{undefined}, \quad \quad  \text{if}\ |N_{\epsilon}(p)| < \text{\textit{MinPts}} \\
\text{max}(\text{core-dist}_{\epsilon , \text{\textit{MinPts}}}(p), dist(p,o)), & \text{otherwise.} \
\end{cases}
\end{alignat}
\end{small}

\noindent
Fig.~\ref{fig:core2} shows examples of reachability distances for some points.

\end{itemize}

\begin{figure}[h!]
\includegraphics[width=\linewidth]{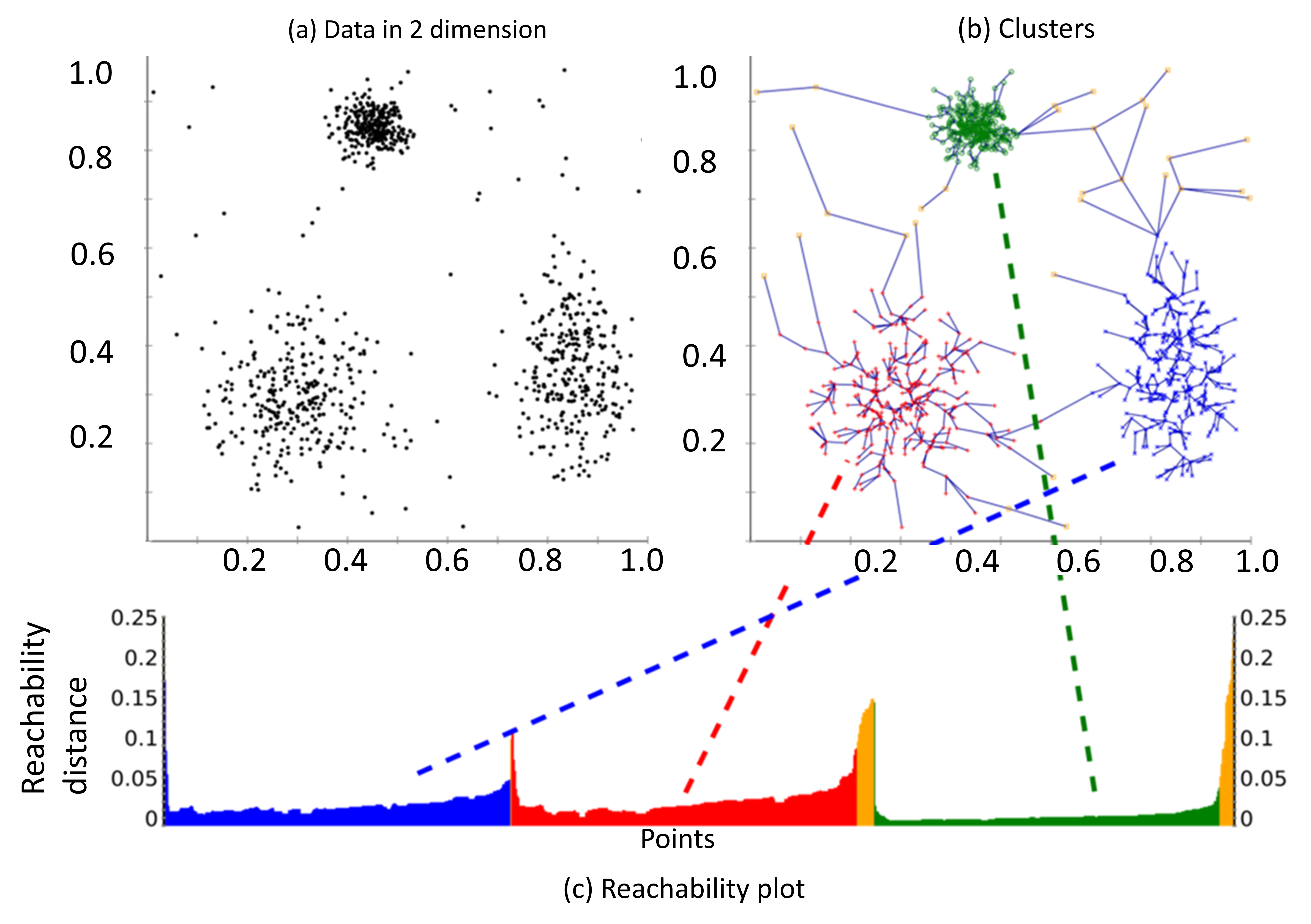}
\caption{OPTICS: (a) Sample data in 2 dimension, (b) Clusters, (c) Reachability plot for all points in the sample data. Fig. source. https://github.com/christianversloot/machine-learning-articles/blob/main/performing-optics-clustering-with-python-and-scikit-learn.md}
\label{fig:reach}
\end{figure}

In OPTICS clustering, first, core distances of all points are computed. Then, we calculate their reachability distances  iteratively. In this iterative process, the data points are chosen in ascending order of their lowest reachability distance to the points already visited until that time. Finally, based upon the reachability distances of all points, we construct a reachability plot as in Fig.~\ref{fig:reach}(c). 

 In Fig.~\ref{fig:reach}(c), we observe valleys of points (sequence of low reachability distances sandwiched between high reachability distances), which represent high density regions. Valleys get separated by regions of low-density. These valleys can be considered to be part of the same cluster(Fig.~\ref{fig:reach}(b). For example, the colored points(other than gray) are those identified as clusters, while gray ones represent noise. For more details on OPTICS, we refer to the original paper~\cite{ankerst1999optics}.

 \section*{S4. Parameter details of GraphSage}
\label{app:param}
Hyper-parameters used for training the GraphSage \gnn{} model are given below. Note that these parameters were optimized based on the results on the training and validation data.

\begin{itemize}
    \item Number of \gnn{} Layers($k$) = 2
    \item Hidden layer size = 64
    \item Random Walk Length($L$) = 4
    \item No. of Random Walks per node = 10
    \item Number of Negative Samples ($Q$) = 10
    \item Optimizer = Adam
    \item Learning rate = 0.01
    \item Graph batch size = 5
    \item Node batch size = 250
    \item Neighborhood aggregation function = Mean
    \item Gradient clipping max norm = 5
    
\end{itemize}

\begin{figure}[h!]
\includegraphics[width=0.5\linewidth]{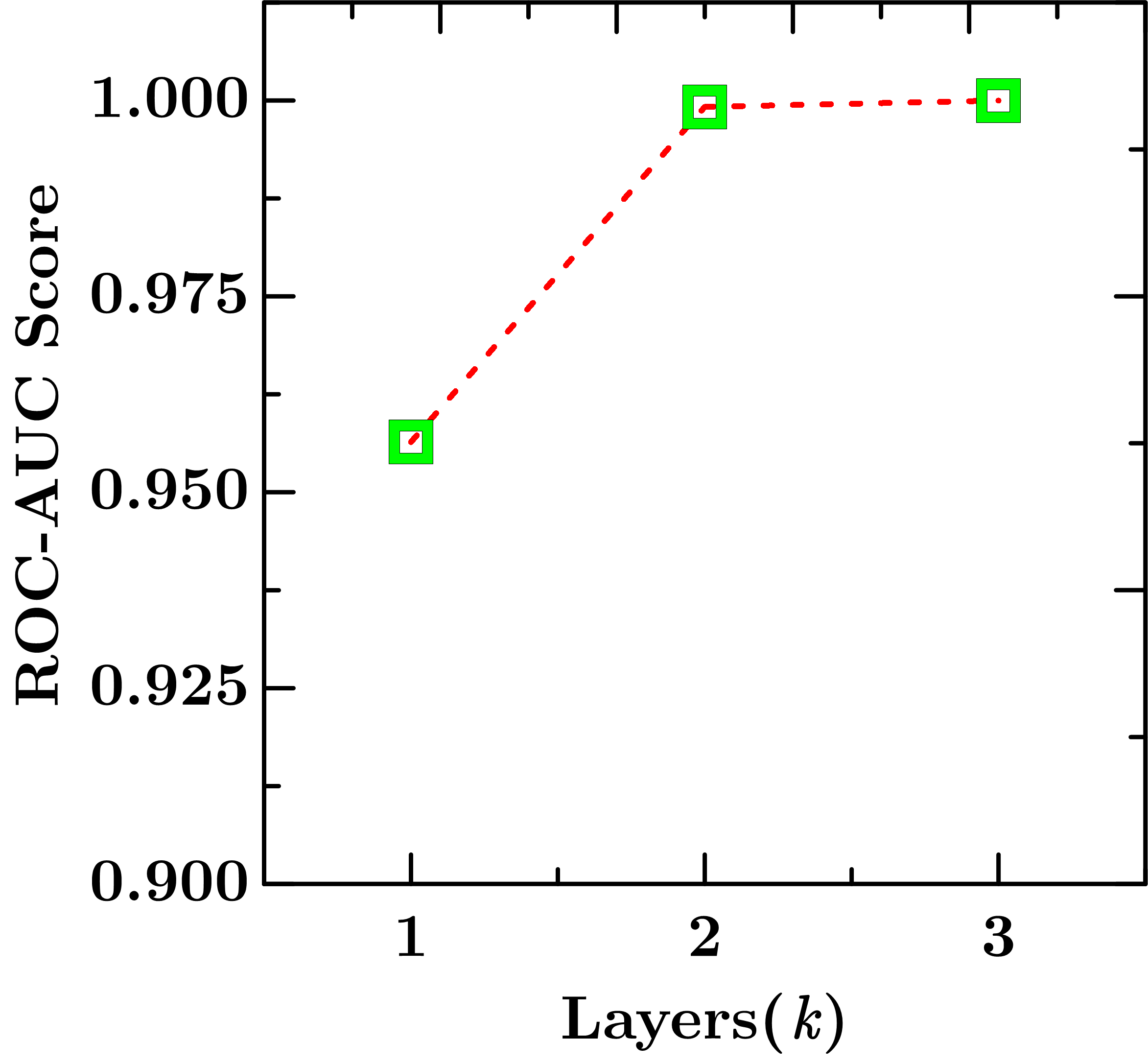}
\caption{ROC-AUC score at different number of \gnn{} Layers($k$) for random walk test}
\label{fig:Layers_test}
\end{figure}

\textbf{Choice of number of layers ($k$)}: As discussed in the main paper, each node aggregates information from its $k$-hop neighborhood. To identify a suitable value of $k$, we evaluate the model by performing a random walk test on the set of validation graphs (not seen during training). As per our training loss in Eq.~\ref{gnn:loss}, the cosine similarity of embeddings of a pair of nodes co-occurring in random walks is expected to be higher than cosine-similarity of embeddings of node pairs that do not co-occur in random walk. In Fig.~\ref{fig:Layers_test}, we show the Area under cover- Receiver Operating characteristics (AUC-ROC) score obtained from random walk test for \gnn{} model trained at different number of \gnn{} layers ($k$). We observe that the performance of the model at $k=2$ is significantly higher than at $k=1$. Further, the performance saturates at $k=2$. Hence, we use $k=2$, which provides a good balance between efficacy and efficiency (computation cost grows with the number of layers).

\section*{S5. Parameter details of OPTICS}
We used OPTICS clustering from Scikit-learn library\cite{opticsscikit}. We mention the parameters used below:

\begin{itemize}
    \item $MinPts$: Minimum number of samples in a neighborhood for a point to be considered as a core point. We set this to $5$.
    
    \item $\epsilon$: Maximum distance between two samples for one to be considered as in the neighborhood of the other. We set this to $5$.
    
    \item Distance metric: Cosine similarity
    
\end{itemize}

\section*{S6. Reachability plot for nodes of 2D LJ System}

 \begin{figure}[h!]
\includegraphics[width=\linewidth]{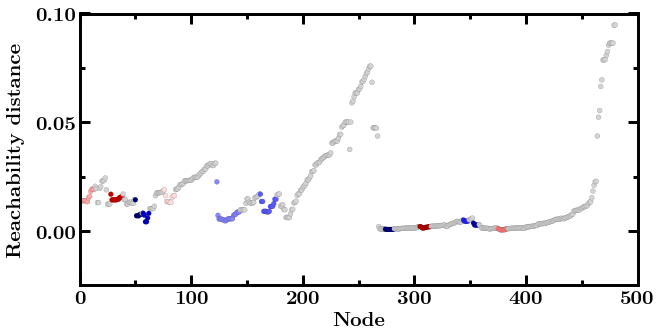}
\caption{Reachability distance plot generated by  OPTICS for nodes of  2D LJ system}
\label{fig:ourreach}
\end{figure}

In Fig.~\ref{fig:ourreach} we show the reachability distance plot obtained by OPTICS algorithm for nodes of our 2D LJ system (same as shown in Fig.~\ref{fig:cluster}.b in main paper). Each  color represents a cluster generated by OPTICS as in Fig.~\ref{fig:cluster}.b. The gray colored points(noise) instead of forming a cluster themselves, act as a separator between regions of high density.

\section*{S7. Effect of random walk length}

We also study the effect of length of random walk in  Table~\ref{tab:pearson}. We observe that the correlation between MSD of clusters and average cluster potential energy increases monotonically with the length of random walk till $4$, after which it drops. 
A possible reason is that when the length of the random walk increases beyond $4$, far away nodes could get incorporated in the set of \textit{positive samples} pairs, and the model might accumulate noise from these node pairs at larger distances. 

\begin{table}[h!]
\centering
\begin{tabular}{c c}
\hline
{Random Walk Length} & Correlation \\
{2} &{0.658}\\
 {3} &{0.714}\\ 
{4} & {0.747}\\
{5} & {0.627} \\
\hline
\end{tabular}
\caption{ \label{tab:pearson}  Pearson’s correlation coefficient between MSD and average cluster potential  energy}
\end{table}